\documentclass[a4paper,11pt]{article}
\pdfoutput=1 

\usepackage{jcappub} 

\usepackage[T1]{fontenc} 

\title{Absorption of massless scalar field by furry black holes in de Rham-Gabadadze-Tolley theory}

\author[a]{Ping Li,}
\author[b]{Yang Huang,}
\author[a]{Xin-Zhou Li}

\affiliation[a]{Division of Mathematical and Theoretical Physics, Shanghai Normal University, 100 Guilin Road, Shanghai 200234, China}
\affiliation[b]{School of Physics and Technology, University of Jinan, 336 West Road of Nan Xinzhuang, Jinan, Shandong 250022, China}

\emailAdd{Lip57120@shnu.edu.cn}
\emailAdd{sps\_huangy@ujn.edu.cn}
\emailAdd{kychz@shnu.edu.cn}

\abstract{We study the absorption of massless scalar field by two types furry charged black holes in de Rham-Gabadadze-Tolley (dRGT) theory. The absorption cross section is calculated in high frequency limit $\sigma_{hf}$ and low frequency limit $\sigma_{lf}$. We show that the high frequency limit $\sigma_{hf}$ is the area of shadow and the low frequency limit $\sigma_{lf}$ is the area of horizon.
 The ratio $R_{f}=\frac{\sigma_{hf}}{\sigma_{lf}}$ is used to measure the impact of charges on the absorption spectra of furry black hole. If the black hole possess an extra charge except mass, the interval value of absorption ratio $R_{f}$ is different: $[1,\frac{27}{16}]$ for electric charge, $[0.7675,\frac{27}{16}]$ for positive charge and $[\frac{27}{16},3.1835]$ for negative charge. We also use a numerical method to compute the absorption cross section in the finite frequency domain. A series of numerical results are presented.}

\begin{document}
\maketitle
\flushbottom
\section{Introduction}

General relativity (GR) is a uniquely theory of massless spin-2 in four dimensions. Meanwhile, dRGT theory \cite{deRham2011} may be the theory of massive spin-2. A comprehensive review of massive gravity can be found in \cite{deRham2014}.

Similar to GR, the spherically symmetric vacuum solution plays a crucial role in massive gravity. However, there is no conventional Schwarzschild solution in the unitary gauge \cite{Berezhiani2012}. In order to restore the diffeomorphism invariance, the St\"{u}ckelberg fields $\phi^a$ are introduced in dRGT theory. Compared with the process to seek the spherically symmetric vacuum solution in GR, there existed an additional invariant $I^{ab}=g^{\mu\nu}\partial_{\mu} \phi^{a} \partial_{\nu} \phi^{b}$ in massive gravity \cite{Berezhiani2012}. In the unitary gauge $\phi^a=x^{\mu}\delta^a_{\mu}$, any inverse metric $g^{\mu\nu}$ that has divergence including the coordinate singularity in GR would exhibit a singularity in the invariant $I^{ab}$. Then, we use a nonunitary gauge to seek healthy spherically symmetric vacuum solutions. In the Ref.\cite{Li2016}, we obtain seven solutions including the Schwarzschild solution, Reissner-Nordstr$\ddot{o}$m (RN) solution, and furry black holes in dRGT theory. All these solutions have no coordinate singularity in $I^{ab}$. Therefore, these solutions may become candidates for black holes in dRGT.

The spherically symmetric black holes in GR depend on two parameters: mass $M$ and electric charge $Q$. There are numerous papers \cite{Matzner}-\cite{Andersson} and books \cite{Book} have studied the scattering problem in these black holes. For recently work, Dolan \emph{et al.} studied Fermi field scattering by Schwarzschild black hole in Ref.\cite{Dolan2006}. Crispino \emph{et al.} have studied the electromagnetic absorption of RN black hole \cite{Crispino2008}. The absorption of massive scalar field by RN black hole is studied detailed in Ref.\cite{Benone2014}.

Furry black holes possess an event horizon that depends on the Schwarzschild radius $r_s$, electric charge $Q$ and a new parameter $S$. The new parameter $S$ introduce a new degree of freedom in event horizon of black hole, which is not the Mass $M$ and Electric Charge $Q$. Thus, we call this furry black hole. This may be a new classification of black hole beyond Kerr-Newman family.

The furry black holes have some features different from black hole in Kerr-Newman family.
In Ref.\cite{Li2020}, we have shown the supperradiant instabilities for furry black hole in a cavity depend on the black hole's parameters. As a comparison, the supperradiant instabilities for RN black hole in a cavity is free of $M$ and $Q$, which is shown by Degollado \emph{et al.} \cite{Degollado2013}. In this paper, we study the absorption of neutral massless scalar field by two types furry black holes in detail. We also show a clear difference between the absorption spectra of black holes in GR and the absorption spectra of furry black holes in massive gravity. To illustrate this difference, we define the ratio $R_f=\frac{\sigma_{hf}}{\sigma_{lf}}$, whose value can reflect the degree of a curved background by black hole. The typical value $R_f$ for the absorption of massless field by Schwarzschild black hole is $\frac{27}{16}$. In the case of RN black hole, the absorption ratio $R_f$ is the range $[1,\frac{27}{16}]$. In the case of type (I) furry black hole, the ratio $R_f$ has a maximum $\frac{27}{16}$ in Schwarzschild case, the electric charge $Q$ can reduce this value to 1 at most; while the positive charge $k_{+}$ can reduce this value to 0.7675 at most. In the case of type (II) furry black hole, for a fixed $Q$, the absorption ratio $R_f$ has a maximum when $k_{-}$ is the minimum and the absorption ratio $R_f$ has a minimum when $k_{-}$ is the maximum. At intermediate frequencies, the numerical method is used to compute the absorption cross section. In the end, the full spectrum description of absorption is obtained together with the different methods.

This paper is organized as follows. In Sec.II, we derive the expression of the absorption cross section by two types furry black holes. In Sec.III and Sec.IV, we calculate the absorption cross section in high and low frequency limit. In Sec.V, we numerical calculate the absorption ratio influenced by different charges. In Sec.VI, the numerical method is used to obtain the full spectrum of the absorption cross section. A series of results are presented together with the absorption cross section in high and low frequency limit. Finally, Sec.VII is a conclusion about our result.
\section{Furry black hole and field equations}
The action of dRGT theory is given by
\begin{equation}\label{actiondRGT}
  S=\frac{M_{pl}^2}{2}\int d^4x\sqrt{-g}[R+m^2 U(g,\phi^a)],
\end{equation}
where $R$ is the Ricci scalar of physical metric $g_{\mu\nu}$, $\phi^a$ is the St$\ddot{u}$ckelberg field and $U$ is a potential for the graviton. The potential is composed of three parts,
\begin{equation}\label{potentialmass}
  U(g,\phi^a)=U_2+\alpha_3U_3+\alpha_4U_4,
\end{equation}
where $\alpha_3$ and $\alpha_4$ are dimensionless parameters, and
\begin{eqnarray}
U_2&=&[\mathcal{K}]^2-[\mathcal{K}^2],\\
U_3&=&[\mathcal{K}]^3-3[\mathcal{K}][\mathcal{K}^2]+2[\mathcal{K}^3],\\
U_4&=&[\mathcal{K}]^4-6[\mathcal{K}]^2[\mathcal{K}^2]+8[\mathcal{K}][\mathcal{K}^3]+3[\mathcal{K}^2]^2-6[\mathcal{K}^4].
\end{eqnarray}
Here the square brackets denote the trace, i.e.$[\mathcal{K}]=\mathcal{K}^{\mu}{}_{\mu}$ and
\begin{eqnarray}
\mathcal{K}^{\mu}{}_{\nu}&=&\delta^{\mu}{}_{\nu}-\sqrt{g^{\mu\alpha}\partial_{\alpha}\phi^a\partial_{\nu}\phi^b\eta_{ab}}\nonumber\\
&\equiv&\delta^{\mu}{}_{\nu}-\sqrt{\Sigma}^{\mu}{}_{\nu},\label{K}
\end{eqnarray}
where the matrix square root is $\sqrt{\Sigma}^{\mu}{}_{\alpha}\sqrt{\Sigma}^{\alpha}{}_{\nu}=\Sigma^{\mu}{}_{\nu}$.
Variation the action with respect to the metric leads to the modified Einstein equations
\begin{equation}
  G_{\mu\nu}+m^2T^{(\mathcal{K})}_{\mu\nu}=T^{(m)}_{\mu\nu},
\end{equation}
 where
\begin{equation}
  T^{(\mathcal{K})}_{\mu\nu}=\sqrt{-g}\frac{\delta(\sqrt{-g}U)}{\delta g^{\mu\nu}}.
\end{equation}

In order to obtain the static spherically symmetric black hole, we showed that a self-consistent ans$\ddot{a}$tz should be written as \cite{Li2016}
\begin{eqnarray}
ds^2&=&-b^2(r)dt^2+a^2(r)dr^2+r^2d\Omega^2,\label{sssmetric}\\
\phi^0&=&t+\tilde{h}(r),\\
\phi^i&=&\beta x^i,
\end{eqnarray}
where $a^2(r)=b^{-2}(r)=f(r)$ and $\beta=1$ or $\frac{2}{3\alpha_3}+1$. Actually, we obtain seven solutions including the Schwarzschild solution, Reissner-Nordstr$\ddot{o}$m (RN) solution, and furry black hole solutions in Ref.\cite{Li2016}. Moreover, all of them can avoid the singularity in the invariant $I^{ab}$ from the divergence of $g^{\mu\nu}$ in the unitary gauge. Among the above solutions, the furry black hole excite our interests. For furry black hole with electric charge $Q$, the function $f(r)$ can be written as
\begin{equation}\label{furrymetric}
 f(r)=1-\frac{2M}{r}+\frac{2+\lambda}{2-\lambda}\frac{Q^2}{r^2}-\frac{S}{r^{\lambda}},
\end{equation}
where the cosmological constant term is neglected and $\lambda>2$. As a classic theory, we expect $\lambda$ is an integer. The term $r^{-\lambda}$ , which dubbed to St$\ddot{u}$ckelberg hair, is introduced by St$\ddot{u}$ckelberg field $\phi^a$ in dRGT model. To make the effect of $r^{-\lambda}$ most obvious in large scale $r>>M$, we choose $\lambda=3$. It also has an general meaning for the integer of $\lambda>2$.

We define $H(r)\equiv f(r)r^3$ as the horizon function, where the horizon $r=h$ are the null points of $H(r)$ at $r>0$. Based on rigorous mathematical analysis of function $H(r)$, we actually found two different horizons described by metric (\ref{furrymetric}) in the case $\lambda=3$. When $S$ is in the range $[0,+\infty)$, there existed only one null point $H(r)$ at $r>0$. This case corresponds to black hole with only one horizon similar to Schwarzschild. When $S\in(-\frac{2M^3}{27}g_{+},0)$, where $g_{\pm}\equiv8+45\frac{Q^2}{M^2}\pm(4+15\frac{Q^2}{M^2})^{\frac{3}{2}}$, there are two null points of $H(r)$ at $r>0$. This case corresponds to black hole with two horizons similar to RN \cite{Li2020}. When $S<-\frac{2M^3}{27}g_{+}$, there are no null points of $H(r)$ at $r>0$, which corresponds to a naked singularity.

As we have known, for black holes in Kerr family, their horizon(s) are decided by conservation charge. For a set of conservation charge of black holes, it decides one and only one type of horizon. If a set of conservation charge have more than one type horizon, then conservation charge cannot decide all properties of black hole. It means that in order to distinguish which one is practical, a nonphysical method is needed. We don't think this is physical. Thus, for black holes beyond Kerr family, we can also expected that a set of conservation charge decides one and only one type of horizon.

According to the horizon(s) of $H(r)$, we can define two different types of conservation charge as follow:
\begin{enumerate}
  \item Positive Charge $k_{+}$: The furry black hole has only one horizon $h$ which is the zero point of function $H(r)$ in the range $r>0$. Thus, function $H(r)$ can be factorized to $H(r)=(r-h)(r^2+a r+b)$.
      In order to recover (\ref{furrymetric}), we can relate parameter $S$ as the set of conservation charge $(M,Q,k_{+})$
      \begin{eqnarray}
       S&=&(M+\sqrt{M^2+5Q^2+k_{+}^2})k_{+}^2\nonumber\\
       &\equiv&hk_{+}^2,
       \end{eqnarray}
       where $h$ is the horizon of furry black hole. Then the metric function $f(r)$ can be factorized to
       \begin{equation}\label{CASEI}
       f_{I}(r)=\frac{(r-h)(r^2+(h-2M)r+k_{+}^2)}{r^3}.
       \end{equation}
      This actually happens for $S>0$ and the range of $k_{+}$ is $[0,\infty)$. We call this type (I) black hole.

  \item Negative Charge $k_{-}$: The furry black hole have two horizons $h_1$ and $h_2$ similar to RN black hole. And function $H(r)$ can be factorized to $H(r)=(r-h_1)(r-h_2)(r+a)$. Similar to type (I) black hole, we can define parameter $S$ as the set of conservation charge $(M,Q,k_{-})$
      \begin{equation}
       S=k_{-}h_1h_2,
      \end{equation}
      where
     \begin{eqnarray}
      h_1&=&\frac{1}{2}(2M-k_{-}-\sqrt{4M^2+20Q^2+4Mk_{-}-3k_{-}^2}),\\
      h_2&=&\frac{1}{2}(2M-k_{-}+\sqrt{4M^2+20Q^2+4Mk_{-}-3k_{-}^2})
      \end{eqnarray}
      are two horizons of furry black hole. The metric function $f(r)$ now can be factorized to
       \begin{equation}\label{CASEII}
       f_{II}(r)=\frac{(r-h_1)(r-h_2)(r-k_{-})}{r^3}.
       \end{equation}
      This actually happens for $S\in(-\frac{2M^3}{27}g_{+},0)$ and we call this type (II) black hole. In order to satisfy $S\in(-\frac{2M^3}{27}g_{+},0)$, the range of $k_{-}$ is $(\frac{2}{3}(M-\sqrt{15Q^2+4M^2}),M-\sqrt{M^2+5Q^2})$. Notice that $k_{-}$ is always negative, thus we call $k_{-}$ negative charge.
\end{enumerate}
The strength of St$\ddot{u}$ckelberg hair is now described by $k_{\pm}$ which can affect the size of horizon $h$. Actually, the configuration of St\"{u}ckelberg field led to new feature of black hole in dRGT theory. Thus, both Positive Charge $k_{+}$ and Negative Charge $k_{-}$ are known as St$\ddot{u}$ckelberg charge.

In the following part, we will calculate the absorption of neutral massless scalar field by two types black holes respectively. The Klein-Gordon equation for scalar fields in such a curved background can be written as
\begin{equation}\label{KG}
\nabla_{\mu}\nabla^{\mu}\Phi=0.
\end{equation}
Assume that the incident wave is along the $z$-axis, thus the complex scalar field can be separated as
\begin{equation}\label{ansatz}
\Phi_{\omega l}=\frac{\psi_{\omega l}(r)}{r}P_l(\cos\theta)e^{-i\omega t},
\end{equation}
where $P_l(\cos\theta)$ is a Legendre polynomial. Plugging the metric (\ref{sssmetric}) into evolution equation (\ref{KG}), we obtain the radial equation
\begin{equation}\label{radial}
f^2\psi''_{\omega l}+ff'\psi'_{\omega l}+\big[\omega^2-f(\frac{f'}{r}+\frac{l(l+1)}{r^2}) \big]\psi_{\omega l}=0.
\end{equation}
The tortoise coordinate $r^*$ is defined as $dr^*=\frac{dr}{f(r)}$. In both types (\ref{CASEI}) and (\ref{CASEII}), we obtain
\begin{eqnarray}
 r_{I}^{*}&=&r+\beta_1\frac{\arctan\big[\frac{2r+h-2}{\sqrt{4k_{+}^2-(h-2)^2}} \big]}{\sqrt{4k_{+}^2-(h-2)^2}}+\beta_2\ln[r-h]+\beta_3\ln[r^2+(h-2)r+k_{+}^2],\\
 r_{II}^{*}&=&r+\frac{h_1^3\ln(r-h_1)}{(h_1-h_2)(h_1-k_{-})}+\frac{h_2^3\ln(r-h_2)}{(h_2-h_1)(h_2-k_{-})}-\frac{k^3\ln(r-k_{-})}{(h_1-k_{-})(h_2-k_{-})},\label{torii}
\end{eqnarray}
where
\begin{eqnarray}
\beta_1&=&\frac{h(h-2)^3+2(2+h-h^2)k_{+}^2-2k_{+}^4}{k_{+}^2+2h(h-1)},\\
\beta_2&=&\frac{h^3}{k_{+}^2+2h(h-1)},\\
\beta_3&=&-\frac{h(h-2)^2-2k_{+}^2}{2[k_{+}^2+2h(h-1)]},
\end{eqnarray}
and we have chosen $M=1$. By using the tortoise coordinate $r^*$, the radial equation (\ref{radial}) reduce to a schr$\ddot{o}$dinger-like form
\begin{equation}\label{schrodinger}
  \frac{d^2}{dr^{*2}}\psi_{\omega l}(r)+\big[\omega^2-V_0(r)\big]\psi_{\omega l}(r)=0,
\end{equation}
where
\begin{equation}\label{potential}
V_0(r)=f(r)\big(\frac{f'(r)}{r}+\frac{l(l+1)}{r^2}\big),
\end{equation}
is the effective potential.

Taking the asymptotic limits of Eq.(\ref{schrodinger}), we obtain the solutions approaching the boundary
\begin{equation}\label{boundary}
  \psi_{\omega l}(r)\approx\left\{
                             \begin{array}{ll}
                               T_{\omega l}e^{-i\omega r^{*}}, & \hbox{for $r\rightarrow h$,} \\
                               e^{-i\omega  r }+R_{\omega l}e^{i\omega  r}, & \hbox{for $r\rightarrow\infty$,}
                             \end{array}
                           \right.
\end{equation}
where $|R_{\omega l}|^2$ and $|T_{\omega l}|^2$ are interpreted as the reflection and transmission coefficients respectively. The conservation of flux indicate a relationship between them
\begin{equation}\label{conservation}
  |R_{\omega l}|^2+|T_{\omega l}|^2=1.
\end{equation}

The absorption cross section is the ratio of the flux in $\Phi$ passing into the black hole to the current in the incident wave. The total absorption cross section $\sigma$ is a sum of partial cross sections $\sigma_l$
\begin{equation}
  \sigma=\sum_{l=0}^{\infty}\sigma_l,
\end{equation}
where $\sigma_l$ is defined by the transmission/reflection coefficients
\begin{equation}\label{sigmal}
  \sigma_l=\frac{\pi}{\omega^2}(2l+1)(1-|R_{\omega l}|^2)=\frac{\pi(2l+1)}{\omega^2}|T_{\omega l}|^2.
\end{equation}

\section{High-Frequency Regime}
\begin{figure*}
\centering
\includegraphics[width=.4\textwidth]{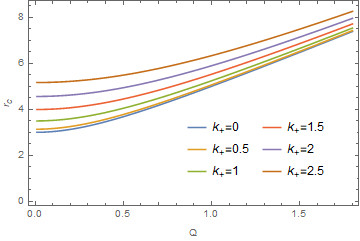}
\includegraphics[width=.4\textwidth]{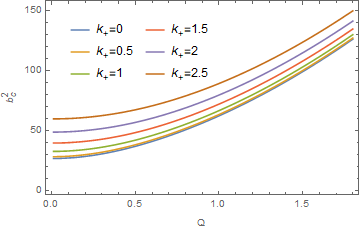}
\caption{ The critical radius $r_c$ and critical impact parameter $b_c^2$ as the function of $Q$ for type (I) black hole. }
\label{Fig:Qv1}
\end{figure*}

\subsection{theoretical analysis}
As the frequency getting higher and higher, the particle properties become more and more apparent. When the wavelength of the field becomes very small in comparison to the size of horizon, the wavefront propagates along geodesics of background. Thus, the absorption cross section is going to approaching the geodesic capture cross section. Without loss of generality, we consider the plane motion for a free particle in $\theta=\frac{\pi}{2}$. Through Killing vector $K^{\mu}$, we can define two conserved quantity: energy $E$ and angular momentum $L$
\begin{eqnarray}\label{conservedq}
E&=&f(r)\frac{dt}{d\lambda},\\
L&=&r^2\frac{d\phi}{d\lambda},
\end{eqnarray}
where $\lambda$ is the affine parameter. Along the geodesic, the mass of particle is also a conserved quantity
\begin{equation}\label{mass}
 0=\mu^2=g_{\mu\nu}\frac{dx^{\mu}}{d\lambda}\frac{dx^{\nu}}{d\lambda}.
\end{equation}
Using the expression for $E$ and $L$, we obtain
\begin{equation}
  -E^2+\big(\frac{dr}{d\lambda}\big)^2+f(r)\frac{L^2}{r^2}=0.
\end{equation}
The effective kinetic energy along geodesic is described by $T(r)\equiv\big(\frac{dr}{d\lambda}\big)^2$.

For a free massless particle incoming from infinity toward central black hole, its finial state of motion have three different cases. Case (a) the particle will reach the perihelion $r_p$, and then be reflected to infinity. Case (c) the particle will fall into black hole. (Or saying, the particle is captured by black hole.) There is a critical case between them -- Case (b) the particle will approach an unstable bound $r_c$, but $never$ reach it. The geodesic capture cross section is defined in this critical case. Case (b) must satisfy two conditions at the same time: (i) $\frac{d}{dr}T(r)\mid_{r=r_p}=0$ and (ii) $T(r)\mid_{r=r_p}=0$. Condition (i) indicates that there is an perihelion $r_p$ along geodesic. While condition (ii) indicates that the free particle $never$ reach the perihelion $r_p$. (If the particle has a positive kinetic energy $T>0$ when reaching the perihelion $r_p$, it would keep moving and be reflect to infinity.) In this case, the critical bound $r_p\equiv r_c$ is also known as photon sphere.

Two conditions (i) and (ii) decide two variables: one is the critical bound $r_c$, the other is the critical impact parameter $b_c$. The impact parameter is defined by $b\equiv\frac{L}{E}$. To make the relationship obviously, we introduce the function
\begin{equation}
  \mathcal{T}(r,b)\equiv\frac{T(r)}{L^2}=\frac{1}{b^2}-\frac{f(r)}{r^2}.
\end{equation}
Since $L$ is constant, the two conditions turn into (i) $\mathcal{T}(r_c,b_c)=0$ and (ii) $\partial_r\mathcal{T}(r_c,b_c)=0$. For the solution $(r_c,b_c)$ of condition (i) and (ii), all geodesics have no perihelion in the range $r<r_c$; meanwhile, all geodesics with parameter $b<b_c$ are captured by black hole. Thus, the geodesic capture cross section is defined by $\sigma_{hf}=\pi b_c^2$. In fact, the critical impact parameter is also the radius of shadow. Thus, the high frequency limit is also the area of shadow from a far view.

\begin{figure*}
\centering
\includegraphics[width=.4\textwidth]{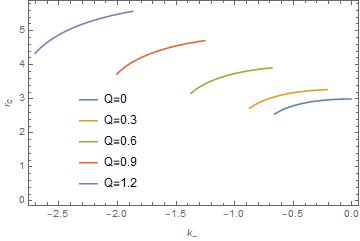}
\includegraphics[width=.4\textwidth]{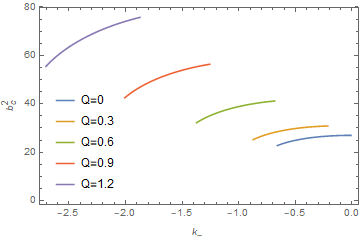}
\caption{ The critical radius $r_c$ and critical impact parameter $b_c^2$ as the function of $k_{-}$ for type (II) black hole. }
\label{Fig:kv1}
\end{figure*}
\subsection{type (I) black hole}
We seek the solutions of conditions (i) and (ii) by Wolfram Mathematica soft. In the actual calculation, there may be multiple sets of solutions. A rigorous mathematical analysis shows that there existed one and only one solutions $(r_c,b_c^2)$ in the range $r>h$. Actually, we only need the set in which both of element is largest. As a common sense, the mass of black hole is chosen to be 1. When all parameters are fixed, the equations of conditions (i) and (ii) are actually polynomial equations. Thus, we use the NSolve method to seek the solutions. To get rid of unwanted solutions, we add the conditions $r>h$ and $reals$ when doing numerical calculation.

Both the electric charge $Q$ and positive charge $k_{+}$ can influence the high frequency limit $\sigma_{hf}$. For a fixed $Q$, the high frequency limit $\sigma_{hf}$ increases as the positive charge $k_{+}$ growth. A similar behavior would happen when the parameter $k_{+}$ is fixed and the parameter $Q$ increases. Fig.\ref{Fig:Qv1} shows that the critical radius $r_c$ and critical impact parameter $b_c^2$ as the function of $Q$ for various values of $k_{+}$. In the Schwarzschild limit $(Q=k_{+}=0)$, the critical radius $r_c=3$ and critical impact parameter $b_c^2=27$, which is exactly shown in Fig.\ref{Fig:Qv1}.

\subsection{type (II) black hole}

In the case of type (II) black hole, the most different thing is the limited impact of $k_{-}$ on the horizon $h_2$. The range of negative charge $k_{-}$ also depends on electric charge $Q$. We plot critical radius $r_c$ and critical impact parameter $b_c^2$ as the function of $k_{-}$ in Fig.\ref{Fig:kv1}, where the critical radius $r_c=3$ and critical impact parameter $b_c^2=27$ when parameters approaching Schwarzschild limit $(Q=k_{-}=0)$.

\section{Low-Frequency Regime}
\subsection{theoretical analysis}
In this section, we are interested in the solutions when wavelength is larger than the horizon of black hole $\omega<<h$. Following Ref.\cite{Unruh1976} and Ref.\cite{Benone2014}, we use a asymptotic expansion technique to solve the equation (\ref{schrodinger}) in various regions. In order to obtain the analytical solution, all terms of higher than $(1/r)^{2}$ are neglected. As a result, the St$\ddot{u}$ckelberg charge $k_{\pm}$ appeared in term $r^{-3}$ is also neglected. However, since $k_{\pm}$ appeared in $h$, the final result still contain the effect of $k_{\pm}$. The most important thing we must emphasize that,
in the case of low frequency, the $l=0$ mode dominant all other $l$ terms \cite{Benone2014}. Therefore, in the following calculation we only concern $l=0$ mode.

We will consider the solution in three different regions: the region near the horizon (Region I), the intermediate region (Region II) and the region far form the black hole (Region III). At last, we match these solutions in different types of black hole.

\begin{enumerate}
  \item Region I: The Eq.(\ref{schrodinger}) is easy to approximate
\begin{equation}\label{LoweqI}
\big( \frac{d^2}{dr^{*2}}+\omega^2\big)\psi=0,
\end{equation}
where $r^*$ is the tortoise coordinate.
  \item Region II: The term of $\omega^2$ is much smaller than all other terms:
\begin{equation}\label{LoweqII}
  \big(\frac{\psi}{r}\big)''+(\frac{2}{r}+\frac{f'}{f})\big(\frac{\psi}{r}\big)'=0.
\end{equation}
  \item Region III: All terms of order  higher than $(1/r)^2$ are neglected in Eq.(\ref{schrodinger})
\begin{equation}\label{LoweqIII}
  (f^{\frac{1}{2}}\psi)''+\big(\omega^2+\frac{4M\omega}{r}
\big)f^{\frac{1}{2}}\psi=0.
\end{equation}
\end{enumerate}

\subsection{type (I) black hole}
For type (I) black hole, it would be more easier when we neglect $r^{-3}$, the tortoise coordinate $r^*$ become
\begin{equation}
  r^{*}_{I,kmin}=r+\frac{h^2}{2(h-M)}\ln(r-h)-\frac{(h-2M)^2}{2(h-M)}\ln(r+h-2M).
\end{equation}
The solution of three different regions are given by
\begin{equation}\label{LowIS}
 \psi=\left\{
        \begin{array}{ll}
          A^{tra}e^{-i\omega r^{*}_{I,kmin}}, & \hbox{for Region I;} \\
          \bigg(\zeta\ln\big(\frac{r-h}{r+h-2M}\big)+\tau\bigg)r, & \hbox{for Region II;} \\
          aF_{0}(\eta,\omega r)+b G_{0}(\eta,\omega r), & \hbox{for Region III.}
        \end{array}
      \right.
\end{equation}
where $F_{l}(\eta,\omega r)$ and $G_{l}(\eta,\omega r)$ are the regular and irregular spherical wave functions respectively and $\eta=-4M\omega$. There are five constants $A^{tra},\zeta,\tau,a,b$ we need to determine in the following matching process.

Let us consider the solution in the overlap of Region I and II. Near the horizon $h$, we only consider the dominant term of solution in Region I
\begin{equation}
 \psi^{I}=A^{tra}(r-h)^{-i\omega \alpha},
\end{equation}
where $\alpha=\frac{h^2}{2(h-M)}$. This term can also be expanded as the series of $\omega$ as
\begin{equation}
 \psi^{I}\approx A^{tra}(1-i\omega \alpha \ln(r-h)),
\end{equation}
where we neglect the terms higher than order $\omega^2$. Meanwhile, taking the limit $r\rightarrow h$ of solution in Region II, we obtain
\begin{equation}
  \psi^{II}\sim h(\zeta\ln(r-h)-\zeta\ln(2h-2M)+\tau).
\end{equation}
Matching these two solutions, we have
\begin{equation}
  \zeta=-\frac{i\omega\alpha}{h}A^{tra},{ }\tau=\frac{(1-i\omega \beta)}{h}A^{tra},
\end{equation}
where $\beta=\alpha\ln(2h-2M)$.

The overlap between Region II and Region III is defined in $r>>h$ but $1>>\omega r$. For $l=0$, the Coulomb wave functions have following form
\begin{equation}
 F_0(\eta,x)=\rho x,{ }G_0(\eta,x)=\frac{1}{\rho},
\end{equation}
where
\begin{equation}
  \rho^{2}=\frac{\eta}{e^{\eta}-1}.
\end{equation}
Thus, the solution in the low frequency limit in Region III reduce to
\begin{equation}
 \psi^{III}=a\rho\omega  r+\frac{b}{\rho}.
\end{equation}
In the asymptotic limit, the solution in Region II can expressed as
\begin{equation}
  \psi^{II}\approx \tau r-\zeta \cdot 2(h-M).
\end{equation}
Thus, matching these solutions, we obtain
\begin{equation}
 a=\frac{1-i\omega\beta}{\rho \omega}A^{tra}, { }b=ih^2\omega \rho A^{tra}.
\end{equation}

The non-normalized incidence coefficient $A^{inc}$ and reflection coefficient $A^{ref}$ are related to $a$ and $b$ by
\begin{equation}
 A^{inc}=\frac{-a+ib}{2i},{ }A^{ref}=\frac{a+ib}{2i}.
\end{equation}
Therefore, the reflection coefficient is given by
\begin{equation}
  |R_{\omega 0}|^2=\bigg|\frac{A^{ref}}{A^{inc}}\bigg|^2=\bigg|\frac{1-h^2\omega^2\rho^2-i\omega\beta}{1+
h^2\omega^2\rho^2-i\omega\beta}\bigg|^2.
\end{equation}
Further considering the low frequency limit $\omega\approx0$ and expand all term as $\omega$, we obtain the first term
\begin{equation}
  \sigma^{I}_{lf}=\mathcal{A},
\end{equation}
where $\mathcal{A}=4\pi h^2$ is the area of black hole.

\subsection{type (II) black hole}
Following a similar calculation process, we can obtain the total absorption cross section in the low frequency limit for type (II) black hole
\begin{equation}
  \sigma^{II}_{lf}=4\pi h_2^2.
\end{equation}
It is easy to check our result. The conclusion is universal that the total absorption cross section of black hole in low frequency limit is the area of event horizon. The total absorption cross section of Schwarzschild in low frequency limit was firstly given by Unruh \cite{Unruh1976}
\begin{equation}
  \sigma_{lf}^{Sch}=\mathcal{A}^{Sch},
\end{equation}
where $\mathcal{A}^{Sch}=4\pi r_s^2$ is the area of Schwarzschild black hole.

\section{The absorption ratio $R_{f}=\frac{\sigma_{hf}}{\sigma_{lf}}$}
No matter the black hole with hair or without hair, the behavior of the frequency limit for the absorption of massless field are very similar. However, an significant difference is still existed between them. In order to discuss this difference, we introduce the absorption ratio
\begin{equation}
 R_{f}=\frac{\sigma_{hf}}{\sigma_{lf}}.
\end{equation}
Different types of conserved charges have different effects on the frequency limit of the absorption cross section. We have shown as the St\"{u}ckelberg charge $k_{\pm}$ increasing in the case of fixed $Q$, both high frequency limit $\sigma_{hf}$ and low frequency limit $\sigma_{lf}$ increase. While, the low frequency limit $\sigma_{lf}$ increases faster. Therefore, the absorption ratio $R_f$ is no longer the same for black hole with different charges.

\begin{figure}
\centering
\includegraphics[width=.4\textwidth]{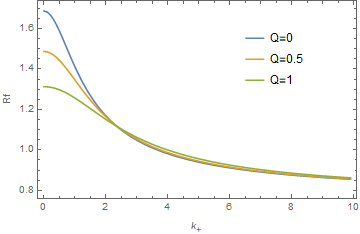}
\caption{ The absorption ratio $R_f$ as the function of $k_{+}$.}
\label{Fig:RfI}
\end{figure}
\subsection{type (I) black hole}
We use a numerical method to calculate $R_f$, which is plotted in Fig.\ref{Fig:RfI}. The lower limit of $R_f$ is only determined by the positive charge $k_{+}$. Whatever the value of electric charge $Q$ is, the lower limit $R_{fmin}=0.7675$ is achieved when the positive charge $k_{+}$ approaching to infinity. The upper limit of $R_f$ is achieved with no positive charge $k_{+}$. The maximum of upper limit is $R_{f}=\frac{27}{16}$ in the Schwarzschild case. If the electric charge $Q$ approaching to infinity, the minimum of upper limit is 1, which is exactly the lower limit $R_{fmin}=1$ decreased by electric charge in the case of RN black hole.

\subsection{type (II) black hole}
\begin{figure}
\centering
\includegraphics[width=.4\textwidth]{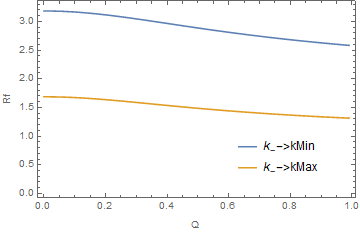}
\caption{ The absorption ratio $R_f$ as the function of $Q$, where $kMin=\frac{2}{3}(1-\sqrt{15Q^2+4})$ and $kMax=1-\sqrt{5Q^2+1}$. }
\label{Fig:RfII}
\end{figure}
In the case of absorption massless field by type (II) black hole, things would be very different. Since the negative charge has a range $k_{-}\in(\frac{2}{3}(1-\sqrt{15Q^2+4}),1-\sqrt{5Q^2+1})$, the lower limit of $R_f$ is achieved when $k_{-}=kMax=1-\sqrt{5Q^2+1}$ and the upper limit of $R_f$ is achieved when $k_{-}=kMin=\frac{2}{3}(1-\sqrt{15Q^2+4})$. The negative charge $k_{-}$ influenced the absorption ratio $R_f$ depend on electric charge $Q$. In the case of no electric charge $Q=0$, the absorption ratio is in the range $R_f\in[\frac{27}{16},3.1835]$. The electric charge $Q$ can also reduce the absorption ratio $R_f$ to 1 in the limit $Q\rightarrow\infty$ and $k_{-}=kMax$.

\section{Numerical Computation}
In this section, we use the numerical method to solve the radial Eq.(\ref{radial}). We expand the solution (\ref{boundary}) near the horizon and approaching infinity. Using Taylor series of solution near the horizon, we integrate the radial Eq.(\ref{radial}) numerically. The integration end at a large $r$. By matching the numerical solution and Taylor series of solution approaching infinity, we obtain the reflection and transmission coefficient.

\subsection{type (I) black hole}
\begin{figure}
\centering
\includegraphics[width=.4\textwidth]{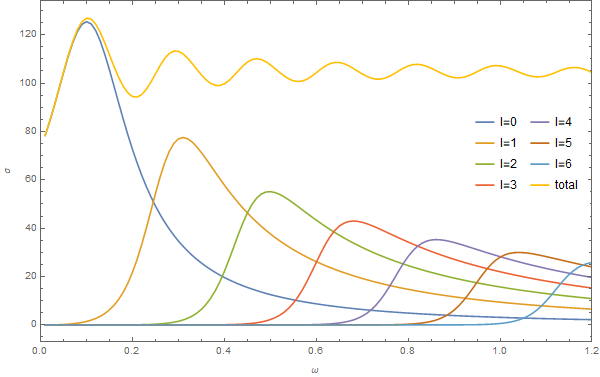}
\caption{ The partial and total absorption cross sections of type (I) black hole, where the parameters are chosen to be $Q=0.2, k_{+}^2=0.8$.}
\label{Fig:IPACS}
\end{figure}

\begin{figure}
\centering
\includegraphics[width=.4\textwidth]{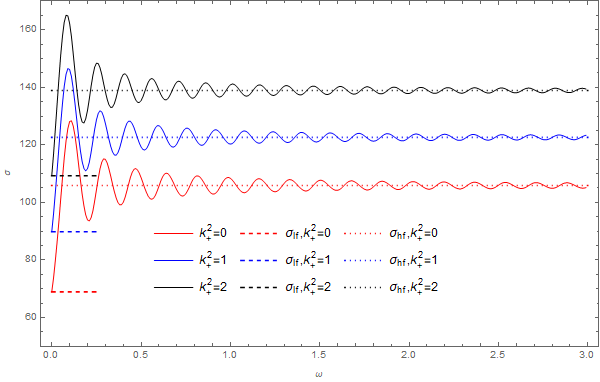}
\caption{ Total absorption cross sections of type (I) black hole in various $k_{+}$. The electric charge is taken to be $Q=0.4$. }
\label{Fig:abIk}
\end{figure}

\begin{figure}
\centering
\includegraphics[width=.4\textwidth]{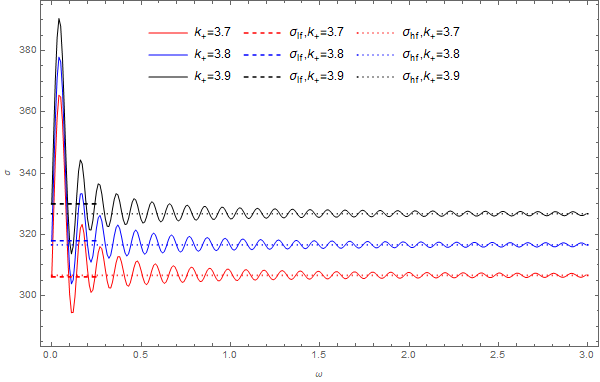}
\caption{ Total absorption cross sections of type (I) black hole, where low frequency $\sigma_{lf}$ limit can exceed high frequency limit $\sigma_{hf}$. The electric charge is also taken to be $Q=0.4$.}
\label{Fig:Trans}
\end{figure}

We present the numerical results for type (I) black hole in this subsection. We calculate partial cross sections $\sigma_{l}$ and sum them to obtain the total absorption cross section, see Fig.\ref{Fig:IPACS}.  Fig.\ref{Fig:abIk} shows the total absorption cross sections $\sigma$ as the function of frequency $\omega$. They fitted well in the high frequency limit and Low frequency limit respectively. Fig.\ref{Fig:Trans} shows the absorption ratio $R_f$ can be smaller than 1. These figures also demonstrate the total absorption cross section $\sigma$ increases as the positive charge $k_{+}$ growth.

\begin{figure}
\centering
\includegraphics[width=.4\textwidth]{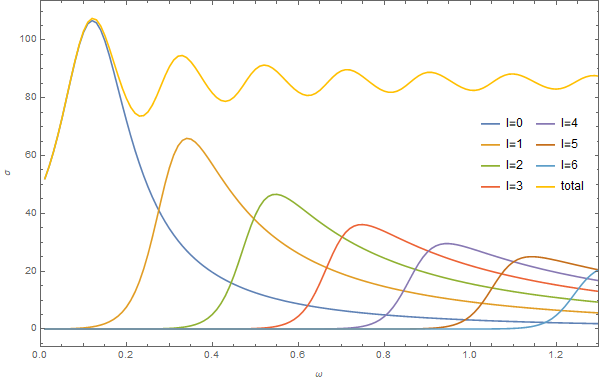}
\caption{ The partial and total absorption cross sections of type (II) black hole, where the parameters are chosen to be $Q=0.2, k_{-}=-0.5$.}
\label{Fig:IIPACS}
\end{figure}

\begin{figure}
\centering
\includegraphics[width=.4\textwidth]{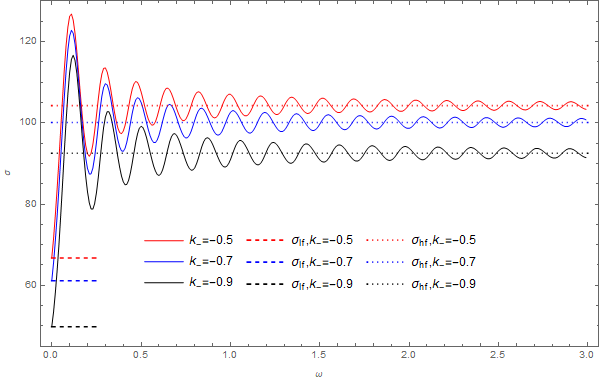}
\caption{ Total absorption cross sections of type (II) black hole in various $k_{-}$. The electric charge is taken to be $Q=0.4$. }
\label{Fig:abIIk}
\end{figure}

\subsection{type (II) black hole}
For type (II) black holes, there are similar numerical results presented in this subsection. In Fig.\ref{Fig:IIPACS}, we show the partial and total absorption cross sections for $Q=0.2$ and $k_{-}=-0.5$.  Fig.\ref{Fig:abIIk} shows the total absorption cross sections $\sigma$ for different values of $k_{-}$. For fixed $Q$, as the negative charge $k_{-}$ decreasing, the total absorption cross section $\sigma$ decreases.

\section{Conclusion}
In this paper, we obtain full spectrum description of absorption spectra of massless field by two types furry black holes. In the high frequency limit, the absorption cross section is the area of shadow. In the low frequency limit, the absorption cross section is the area of horizon.

As a general conclusion, the bigger the black hole is, the more it absorbs the field, which is also shown by the numerical results. The charge $k_{\pm}$, as a new parameter which can affect the horizon, is introduced by the St\"{u}ckelberg field $\phi^a$ in dRGT theory. Thus, the new properties appearing on the absorption cross section is also caused by the configuration of St\"{u}ckelberg field $\phi^a$.

Furthermore, we have shown different charges have a different influence on the absorption ratio $R_f$. If a black hole only has mass $M$, the absorption ratio is a constant $R_f=\frac{27}{16}$. If it also has a electric charge $Q$, as the electric charge $Q$ increasing, the absorption ratio decreases. In the extremely case, the electric charge $Q$ can reduce the absorption ratio to $R_f=1$. The St\"{u}ckelberg hair affects the absorption ratio with similar behavior but totally different value. In the extremely case, the positive charge $k_{+}$ can reduce the absorption ratio to $R_f=0.7675$. While, the negative charge $k_{-}$ can rise the absorption ratio to $R_f=3.1835$ at most.

\end{document}